\documentclass{PoS}

\usepackage{listings}

\title{Clover Action for Blue Gene-Q and Iterative solvers for DWF}

\ShortTitle{BGQ,Iterative solvers for Lattice actions}

\author{\speaker{Karthee Sivalingam}\\
        Department of Meteorology, University of Reading, Reading, UK\\
        E-mail: \email{K.Sivalingam@reading.ac.uk}}

 \author{P.A. Boyle\\
   School of Physics \& Astronomy, University of Edinburgh, EH9 3JZ, UK\\
   E-mail: \email{paboyle@ph.ed.ac.uk}}
   
\abstract{In Lattice QCD, a major challenge in 
simulating physical quarks is the computational complexity of these simulations. 
In this proceeding, we describe the optimisation of Clover fermion action for Blue gene-Q architecture and 
how different iterative solvers behave for Domain Wall Fermion action. We find that 
the optimised Clover term achieved a maximum efficiency of 29.1\% and 20.2\% for single and double precision
respectively for iterative Conjugate Gradient solver. For Domain Wall Fermion action (DWF) 
we found that Modified Conjugate Residual(MCR) as the most efficient solver compared to CG and GCR. 
We have developed a new multi-shift MCR algorithm that is 18.5\% faster compared to multi-shift CG
for the evaluation of rational functions in RHMC.
}

\FullConference{31st International Symposium on Lattice Field Theory - LATTICE 2013\\
		July 29 - August 3, 2013\\
		Mainz, Germany}

\begin{document}

This proceeding describes the optimisation of Clover fermion action for Blue gene-Q architecture and
the application of different iterative solvers for Domain Wall fermion action in two subsequent sections.
 
\section{Clover Action for Blue Gene-Q}

Clover fermion action is widely used in Lattice QCD and is written as 
\begin{equation}
  S_{\rm{clover}}\ =\ S_{W}- \frac{C_{SW}}{4} \sum_{\mu<\nu} \ \bar{\psi}(x)\sigma_{\mu\nu}F_{\mu\nu}\psi(x)\ .
  \label{eq:clover}
\end{equation}
where $S_{W}$ is the Wilson action. Clover action with the right coefficients ($C_{SW}$)
gives ${O}(a)$ improvement for on-shell quantities.
Performance of the inverter is important for any good 
optimisation of Lattice QCD simulation. Inverting sparse fermion matrix involves using an iterative
solver in which clover operator is applied at each iteration.
This section describes the porting and optimisation 
of clover inverter for Blue Gene-Q architecture using the BAGEL compiler ~\cite{Boyle:2009vp}.

Blue Gene-Q is build with the 64-bit Power-PC A2 processor core that has a peak performance 
of 209 Tera flops per rack of 1024 nodes (each node containing 16 compute and one OS core). 
For complete details of the architecture, refer to~\cite{6109225}.
BAGEL is a QCD domain specific library developed by University of 
Edinburgh~\cite{Boyle:2009vp}. Using the BAGEL library, BAGEL Fermion Matrix (BFM) library provides
QCD specific functionality. Currently the library supports solutions to QCD actions Wilson, 
Wilson twisted mass, Domain wall and Overlap. It supports iterative solvers like
Conjugate Gradient (CG), Multi-shift CG in single, double and mixed precisions.

\subsection{Clover and Wilson actions}
Clover action can be written in terms of Wilson action as follows
\begin{eqnarray}
\label{cloverandwilson}
S=\sum_{xy}\bar{\phi(x)}M_{xy}\phi(y) \\ \nonumber
M_{xy}^{\rm{Wilson}}=I-k\,{D}  \\ \nonumber
M_{xy}^{\rm{clover}}=A-k\,{D} \\ \nonumber
A=I-k\,\frac{C_{sw}}{2}\sum_{\mu<\nu}[\gamma_{\mu},\gamma_{\nu}]{F}_{\mu\nu}
\end{eqnarray}
where ${D}$ is Wilson-Dirac operator. For clover action, Wilson-Dirac 
operator ${D}$ and clover term $A$ are applied independently. This clover term $A$
is local and is computed once and then applied to all the iterations of an iterative
solver.
$A$ is hermitian as $[\gamma_{\mu},\gamma_{\nu}]$ and  ${F}_{\mu\nu}$
are hermitian. The algebra of  $\gamma$ matrices leaves A 
with two $6\times6$ hermitian matrices $A_{1}$ and $A_{2}$ at each site.
\begin{equation}
A_{xyzt}=\left(\begin{array}{cc}
A_{1}^{6\times6} & 0\\
0 & A_{2}^{6\times6}\end{array}\right)
\end{equation}
This leaves us with 
implementation of $A\times\phi$ to complete the clover action. In performing
this matrix multiplication, $A_{1}$ and $A_{2}$ 
are represented in a compressed format to save memory. The diagonal elements are 
stored as real numbers and only the lower triangular elements are stored as complex numbers.
\subsection{Optimisation}
BAGEL already has a highly optimised version of 
Wilson-Dirac operator(${D}$)~\cite{boyle2012bluegene}. 
The clover matrix $A$ is constructed 
using an external library like CHROMA or CPS and 
then imported to BAGEL. This leaves us
with optimising only the clover apply kernel($A\times\phi$).  
\subsubsection{SIMD Optimisation}
QPX floating point unit in Blue Gene-Q has a vector length of four. To efficiently use this unit,
data required for four parallel instruction should be aligned consequently in memory.
BAGEL compiler supports aligning data for different vector lengths. The compiler constructs logical SIMD volumes based
on the vector length and stores the data from each of the logical volumes consequently in
memory. With this new data layout, the QPX floating point unit is efficiently used to
increase floating point throughput. 

\subsubsection{Memory Optimisation}
To reduce the memory latency, caches and registers should be efficiently used.
In applying the clover term, all reduction operations should ideally use registers to store 
output variables in order to avoid writing to L1 cache which is write through. With 32 registers available,
it will be easier to load half-spinors into register and compute the results. 
For the efficient implementation of clover apply term, we use 27 registers, 
6 each for $\chi$ and $\psi$ and 15 registers for storing the clover matrix($A$). 

\subsubsection{Instruction pipe-lining}

Io hide memory latency and increase the instruction throughput,
the instructions should be pipelined. 
BAGEL compiler supports constructing two 
pipelines using the greedy algorithm. Dependencies in instructions are identified
and are reordered accordingly. This plan for schedule of instructions is referred 
to as execution map. Execution map is a abstract assembler and this is then translated to 
hardware specific assembly instructions. 
In the execution map, the load(store) instructions are pipe-lined with the multiply
instructions. As the instruction unit is kept busy, this increases 
instruction throughput and the latency
associated with loading(storing) data to memory is hidden.

\begin{figure}[htbp]
\begin{center}
 \includegraphics[scale=0.425, angle=0]{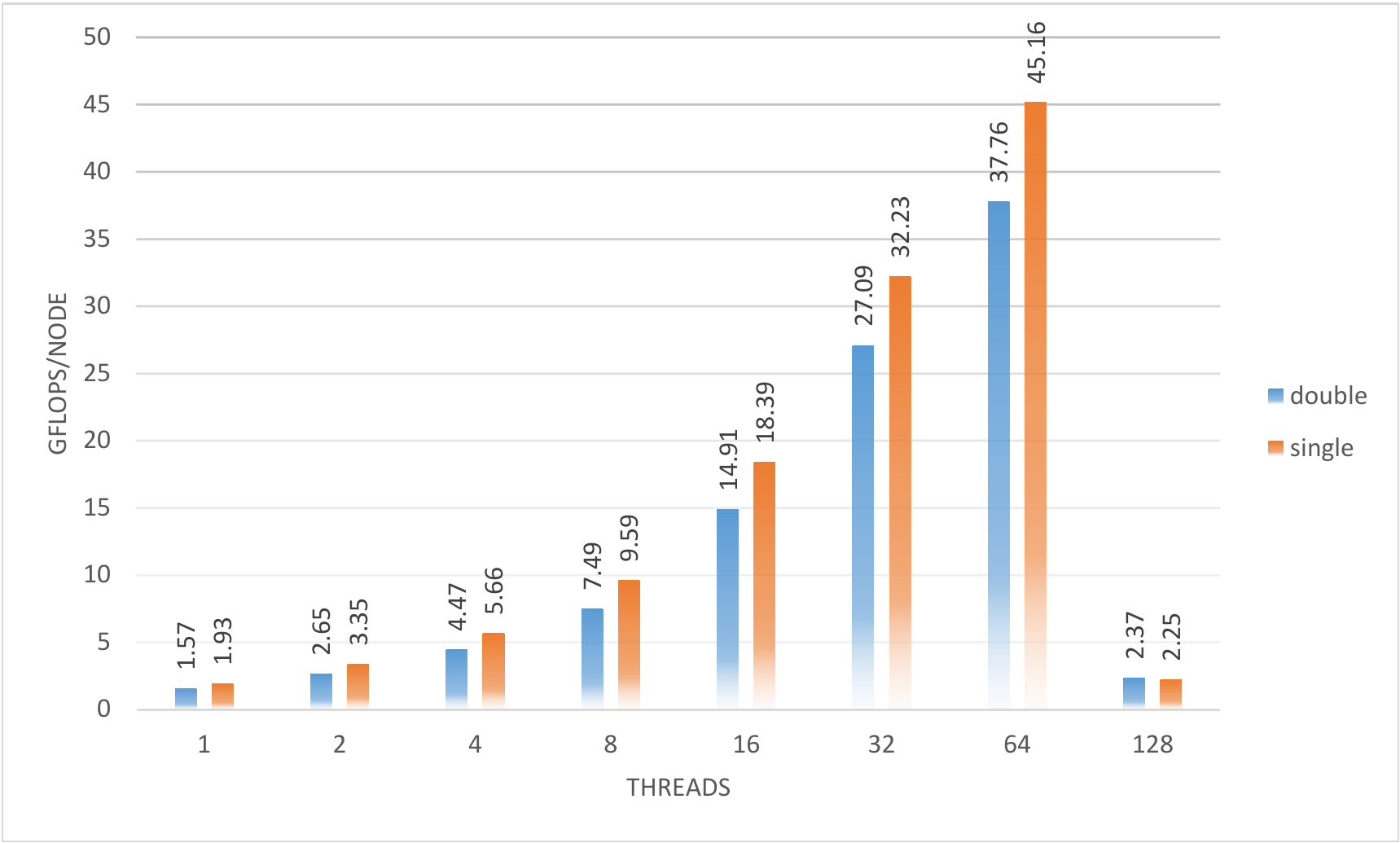}
 \includegraphics[scale=0.286, angle=0]{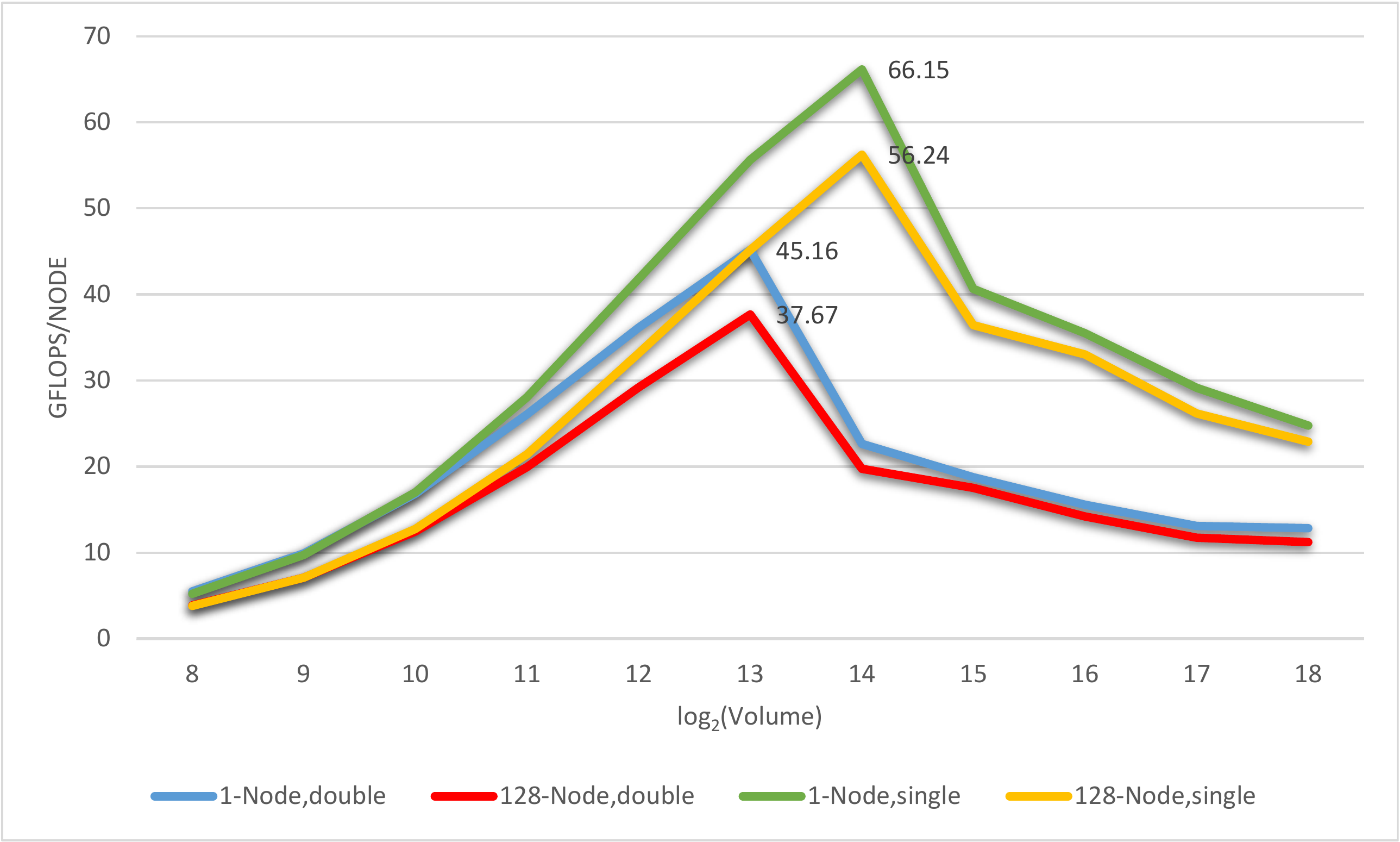}
\caption{Plot showing 
Left : performance in GFlops per node for Clover-CG when
increasing number of threads are used per node. The
performance is measured on lattice volume of $32^4$, running on 128 nodes.
Right : Strong scaling (GFlops per node) of the clover solver in
 double and single precision for increasing local sub-volume (for a single node)
 when run on a single and 128 nodes}
\label{fig-Clover_CG_threads}
\end{center}
\end{figure}

%
\subsection{Results}
The optimisations discussed in previous subsections are applied to clover apply($A$) kernel.
In this section, performance will be measured 
as the performance of the entire iterative Conjugate Gradient solver 
for clover fermion action that
includes application of ${D}$ and $A$ for each iteration. We will simply
refer to it as Clover-CG.
In order to achieve good efficiency, we need to experiment with the threads,
MPI processes and memory. To maximise the usage of
shared memory and reduce unnecessary MPI packets, the application should run only with
one MPI process per node. Each node supports 64 hardware threads and we can experiment with
number of threads that gives optimal performance. 

Fig.~\ref{fig-Clover_CG_threads}(left) shows the performance of Clover-CG in GFlops per node,
for increasing number of threads. The
performance is measured on lattice volume of $32^4$, running on 128 nodes. From the
plot we can infer that we have maximum efficiency of 18\% for double precision
and 23\% for single precision when 64 threads are used. The speedup, when the threads are
increased is not linear. We achieve only
$\approx37\%$ of the expected maximum performance due to the synchronisation overheads.

An important factor in performance for most high performance application
is memory and network bandwidth. Optimally, the data should be available
in cache so that memory latency is reduced. 
For single precision, we achieve a maximum performance of 59.5 GFlops per node when the
lattice volume is $48^4$. And similarly for double precision, 
maximum performance of 41 GFlops per node is achieved for
lattice volume of $32^3 \times 48$. Both the single and double precision performance 
show strong local volume dependence. 

Fig.~\ref{fig-Clover_CG_threads}(right) shows the strong scaling of Clover-CG in
double and single precision for increasing local lattice sub-volume, 
when run on a single and 128 nodes. The strong scaling shows strong dependence on local volume.
This is directly related to the size of the L2cache and maximum performance is
achieved when the data fits the L2cache. This means that to achieve good efficiency we should
run on less or more number of nodes according to the simulated lattice volume.

\section{Iterative solvers for DWF}
Lattice QCD simulations involve computing the Quark propagators in a background gauge fields.
Quark propagators are computed by solving 
\begin{equation}
 ({D}+m_q)\psi(x) = \eta(x)
\label{eq:dirac}
\end{equation}
where ${D}$ is the Dirac matrix, $m_q$ is the quark mass,
 $\psi(x)$ and $\eta(x)$ are
the solution and source field respectively. Iterative methods (see eg.~\cite{saad2003iterative})
are the only viable way to solve large sparse linear systems. In lattice simulation, quark propagators
are computed on different gauge configurations and for different right hand sides. 

For Domain Wall fermion(DWF) action,
the Dirac matrix is large, indefinite and the eigen values
are clustered around the origin. This makes the solution to the linear system 
difficult. Also as the simulated quark masses($m_q$) gets closer to physical values 
and lattice spacing 
($a$) gets smaller, the Dirac matrix becomes ill-conditioned. 
Finding a suitable solver and preconditioner are topics of intense research. 
The following sections discuss iterative solvers namely Conjugate Gradient(CG), 
Generalised Conjugate residual(GCR) and 
Modified Conjugate residual(MCR) (refer ~\cite{saad2003iterative} for details) for 
solving DWF.

\subsection{CG, MCR and GCR }
The iterative methods described in this subsection are Krylov subspace methods
based on projection methods(Petrov-Galerkin conditions). For solving a linear system
$Ax=b$, the Krylov subspace is defined by
\begin{equation}
 {K}_{m}(A,r_0)\equiv span\{r_0,Ar_0,A^{2}r_0,\ldots,A^{m-1}r_0\}
\end{equation}
where $r_0=b-Ax_0$. The approximate solution $x_m$ is obtained by searching in
the subspace $x_0+{K}_{m}$ so that 
\begin{equation}
 b-Ax_{m}\perp{L}_{m}
\end{equation}
where ${L}_{m}$ is also a subspace of dimension m.
Conjugate Gradient (CG) is the most popular 
method for solving sparse symmetric, positive definite linear systems. CG 
uses orthogonal projection (${L}_{m} = {K}_{m}$) on to Krylov 
subspace ${K}_{m}(A,r^0)$. For symmetric, positive definite 
matrices, that are hermitian, MCR improves by
constructing residual vectors that conjugate. 
For non-symmetric matrices,
we can generalise by constructing $p_i$ as a linear combination of 
current and all previous $p_i$s. This general method is referred to 
as Generalised Conjugate Residual.  

For CG, $p_i$s are 
A-orthogonal, whereas for MCR, $Ap_i$s are orthogonal or simply 
$p_i$s are $A^\dagger A$-orthogonal. 
CG and MCR are very similar, but MCR requires storage for one more vector and
requires more operations than CG. GCR algorithm requires us to store all
previous $p_i$s ($Ap_i$s) and this is practically not possible. The number of 
previous $p_i$s that are stored are restricted to a lesser number (m). We can either 
restart after m iterations or truncate the number of $p_i$s stored to the
latest m entries. The former is referred to as GCR(m) and the latter as OrthoMin(m).

\subsection{Results}
In simulating Domain Wall Fermion, 
the fermion matrix is represented as ${M}^{\dagger}{M}$
as it is positive definite and hermitian. In case of GCR and OrthoMin, we can consider both 
${M}^{\dagger}{M}$ and ${M}$, to check if it works generally 
for non-symmetric matrices.  Also for GCR and OrthoMin, careful study is required to balance the 
number of previous residuals to store and computation cost for better performance. 

In this work, we use a variant of CG called CGNE~\cite{Freund92iterativesolution}, which solves
$Ax=b$ by solving $AA^Ty=b$ ($x=A^Ty$). We will refer to it as CG for simplicity.
GCR with fermion matrix ${M}^{\dagger}{M}$ and ${M}$ will be 
referred to as GCR-MM and GCR-M respectively. OrthoMin will also be referred to as O-MIN.
The results described in this section uses gauge configuration with $N_f$= 2+1 dynamical flavors, 
generated from Iwasaki gauge action at $\beta$=2.13 ($a^{-1}$=1.73(4) GeV) and lattice volume 
of $16^3\times32$. All the iterative solvers discussed in this section 
uses $L_s$=16 and quark mass of $0.01$, unless specified otherwise.
The performance is measured on 128 nodes of Blue Gene-Q
machine. 

\begin{figure}[htbp]
\begin{center}
 \includegraphics[scale=0.425, angle=0]{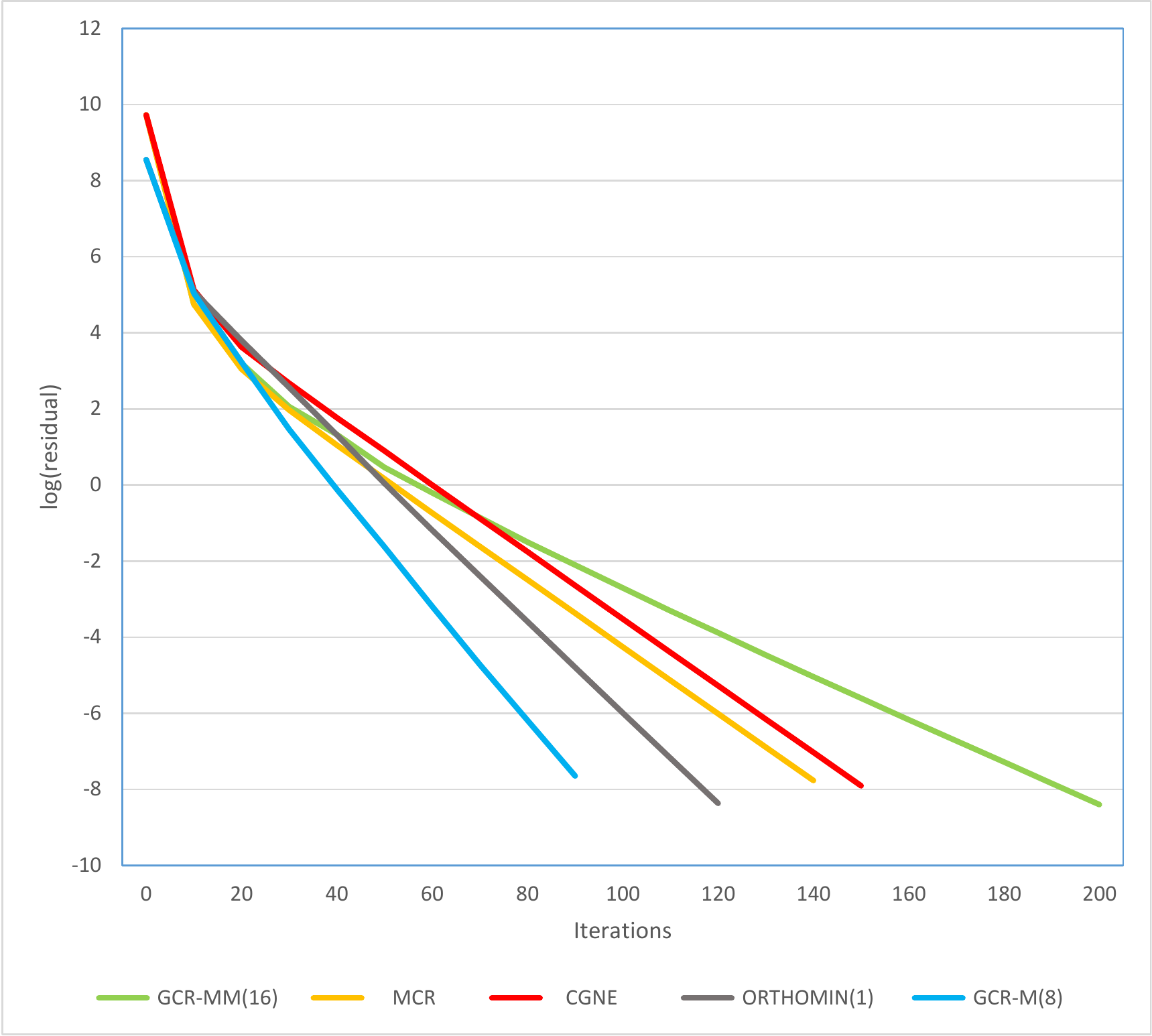}
 \includegraphics[scale=0.39, angle=0]{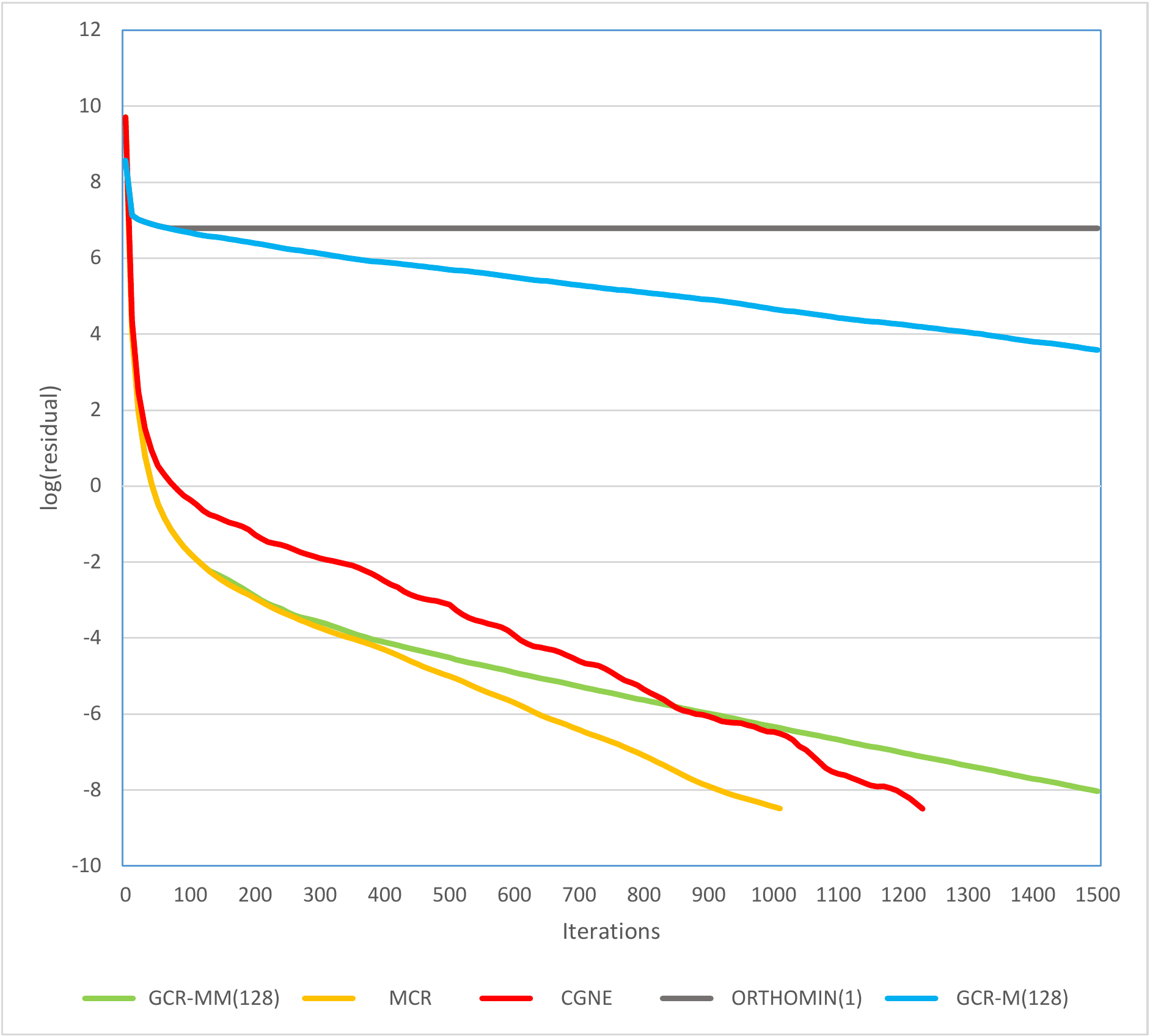}
 \caption{Plot showing how the residual reduces with iterations
for DWF in a random gauge (left) and background QCD gauge field (right), 
with $L_s=16$ using different solvers 
in solving $M^{\dagger}M\,\psi=\chi$. }
\label{fig-dwf_iterations_3000}
\end{center}
\end{figure}

For a random gauge, GCR and OrthoMin
solves ${M}$ efficiently as GCR(4) and OrthoMin(4) solves in almost half the time as that
for CG and MCR.
Fig.~\ref{fig-dwf_iterations_3000} (left) shows convergence of residual as a function of iteration count.
The efficient solvers of GCR and OrthoMin are plotted for reference. The residual 
reduce steeply for GCR, OrthoMin and MCR compared to CG. It is important to note that
where the former methods are based on conjugate residuals, the latter method CG is based
on gradients.

Using a QCD gauge configuration generated using Hybrid Monte-Carlo simulation is interesting as it
changes the spectrum of the DWF Operator. For solving ${M}^{\dagger}{M}\,\psi=\chi$, we see similar 
results as that for random gauge, but the fastest GCR solver is 40 times slower than CG.
For solving non hermitian system ${M}\,\psi=\chi$, GCR and OrthoMin do not converge. A closer study of the 
DWF operator and the impact of the fifth dimension shows that
as $L_s$ increases linearly, the conditioning of the ${M}$ worsens and convergence of GCR
suffers exponentially when compared to CG.
%
%

From numerical analysis in~\cite{Nachtigal:1991,Nachtigal:1992}, we can conclude 
that when the eigen values of the matrix lie in all four quadrants of the 
complex plane, the convergence of non-hermitian solvers(GCR with ${M}$) is unreliable. In 
such cases, normal equations is the best we can do.
Using ${M}^{\dagger}{M}$ is therefore the only option for good convergence.
The GCR and OrthoMin solvers may perform better than the CG, if a good pre-conditioner is used 
as shown by the results from random gauge. Fig.~\ref{fig-dwf_iterations_3000} (right) shows a 
closer look at the convergence of residual as a function of iterations. We can easily identify 
MCR as the most efficient algorithm as it takes 20\% lesser time and number of 
iterations to solve the system.

In solving \ref{eq:dirac}, the solution is usually repeated for different quark masses($m_q$).
Instead of solving them separately, the solution for different quark masses  with same source field
can be computed simultaneously using multi-shift methods~\cite{osborn:2008, Bloch:2009et}. This
is based on the fact that the Krylov subspaces are shift invariant
\begin{equation}
 {K}_{m}({D},b) = {K}_{m}({D}+m,b)
\end{equation}

Multi-shift solvers are a key part in the Rational Hybrid Monte Carlo(RHMC) algorithm.
This method can be used for any of the Kyrlov subspace methods. For DWF, we have found out that
MCR is an efficient algorithm. 
We developed a multi-shift MCR algorithm that uses MCR
as the solver for multiple shifts.
The multiple shifts corresponds to
poles in the rational approximation.
This new multi-shift algorithm accelerates the evaluation of 
rational function by 18.5\% in RHMC algorithm. In 2+1f Lattice simulations, the rational 
function evaluation takes 1/3 of the compute time and using this method will give a overall 
6\% gain in RHMC.
\section{Conclusions}
We have successfully ported the Clover Lattice fermion action to Blue Gene/Q architecture.
  The optimised Clover term achieved a maximum efficiency of 29.1\% and 20.2\% for single and double precision
  respectively for iterative Conjugate Gradient solver.
  This optimised version showed good Weak scaling.
  Strong scaling showed local volume dependency due to the effects of cache capacity and network
  bandwidth. We have studied the different iterative solvers for Domain Wall Fermion action (DWF) and
  found that Modified Conjugate Residual(MCR) as the most efficient solver compared to CG and GCR. 
  We have developed a new multi-shift MCR algorithm that is 18.5\% faster compared to multi-shift CG
  for the evaluation of rational functions in RHMC.

\end{document}